\documentclass[12pt]{iopart}

\usepackage{graphicx}
\usepackage{subfigure}

\newcommand{\eff}{{\mathrm{eff}}}

\begin{document}

\paper{Optical properties of carbon nanofiber photonic crystals}
\author{R. Rehammar$^1$, R. Magnusson$^2$, A. I. Fernandez-Dominguez$^3$, H. Arwin$^2$, J.M. Kinaret$^1$, S. A. Maier$^3$, E.E.B. Campbell$^4$
\address{$^1$ Department of Applied Physics, Chalmers University of Technology, SE-412 96 Gothenburg, SWEDEN}
\address{$^2$ Department of Physics, Chemistry and Biology,  Link\"oping University, SE-581 83 Link\"oping, SWEDEN}
\address{$^3$ Department of Physics, Imperial College London, London SW7 2AZ, United Kingdom}
\address{$^4$ EaStCHEM, School of Chemistry, Edinburgh University, Edinburgh EH9 3JJ,
SCOTLAND and Division of Quantum Phases and Devices, School of Physics, Konkuk University, Seoul 143-701 KOREA}
\ead{robert.rehammar@chalmers.se}}


\begin{abstract}
Carbon nanofibers (CNF) are used as components of planar photonic crystals. Square and rectangular lattices and random patterns of vertically aligned CNF were fabricated and their properties studied using ellipsometry. We show that detailed information such as symmetry directions and the band structure of these novel materials can be extracted from considerations of the polarization state in the specular beam. The refractive index of the individual nanofibers was found to be $n_{CNF}$ = 4.1.
\end{abstract}

\maketitle

Two-dimensional (2D) photonic crystal (PC) slabs are materials with a periodic modulation of the refractive index in a 2D plane and a finite extension in the perpendicular direction. These systems have generated much interest in recent years as a means to control light at length scales on the order of the wavelength \cite{vlasov_mode_2004}.

In this article we report the study of properties of 2D PC slabs consisting of free-standing carbon nanofibers (CNF) grown on a metallic substrate. The optical properties have been investigated using the technique of variable-angle spectroscopic ellipsometry \cite{humlicek_polarized_2005}.

PCs based on CNFs are thought to be promising systems for the development of tunable PCs in the visible range \cite{rehammar_nanowire-based_2008}, making use of the favorable actuation properties of the CNF. In order to fully exploit this potential application, it is important to first understand the static optical properties of such 2D arrays. It was earlier demonstrated that vertical CNF can act as optical antennas to transmit light in a fashion similar to the transmission of radio waves in RF applications \cite{wang_receiving_2004}. Two-dimensional lattices of CNF have also been shown previously to act as diffraction gratings \cite{kempa_photonic_2003, rybczynski_visible_2006}. However, there has been no experimental determination of the band structure of the CNF-based PC and corresponding detailed comparison with theoretical predictions.

The results reported here demonstrate that it is possible to map out large regions of the band structure of the CNF-PC using ellipsometry. The measurements are in good agreement with finite-difference time domain \cite{taflove_computational_2000} band structure calculations using a refractive index of $n_{CNF}$ = 4.1 for the CNF. It is also shown that the symmetry directions of the samples can be easily detected. 

Two-dimensional PC slabs consisting of free-standing CNFs are directly grown on a metallic underlayer using plasma-enhanced chemical vapour deposition (PE-CVD). The technique, with some modifications to the growth parameters, has been described previously \cite{kabir_fabrication_2006}. The lattice structure is defined by using electron-beam lithography to pattern Ni catalyst dots with the required lattice parameters. Arrays were patterned with lattice constants, $a$, ranging between $500$ nm and $200$ nm in different configurations along the $x$- and $y$-directions. Random samples were patterned so that the average CNF density corresponded to that of the square lattices. As an example, for comparison with the $a = 500$ nm lattice constant, CNFs are placed randomly with $0.8$ \% filling factor.

\begin{figure}[!b]
    \includegraphics[width=.95\linewidth]{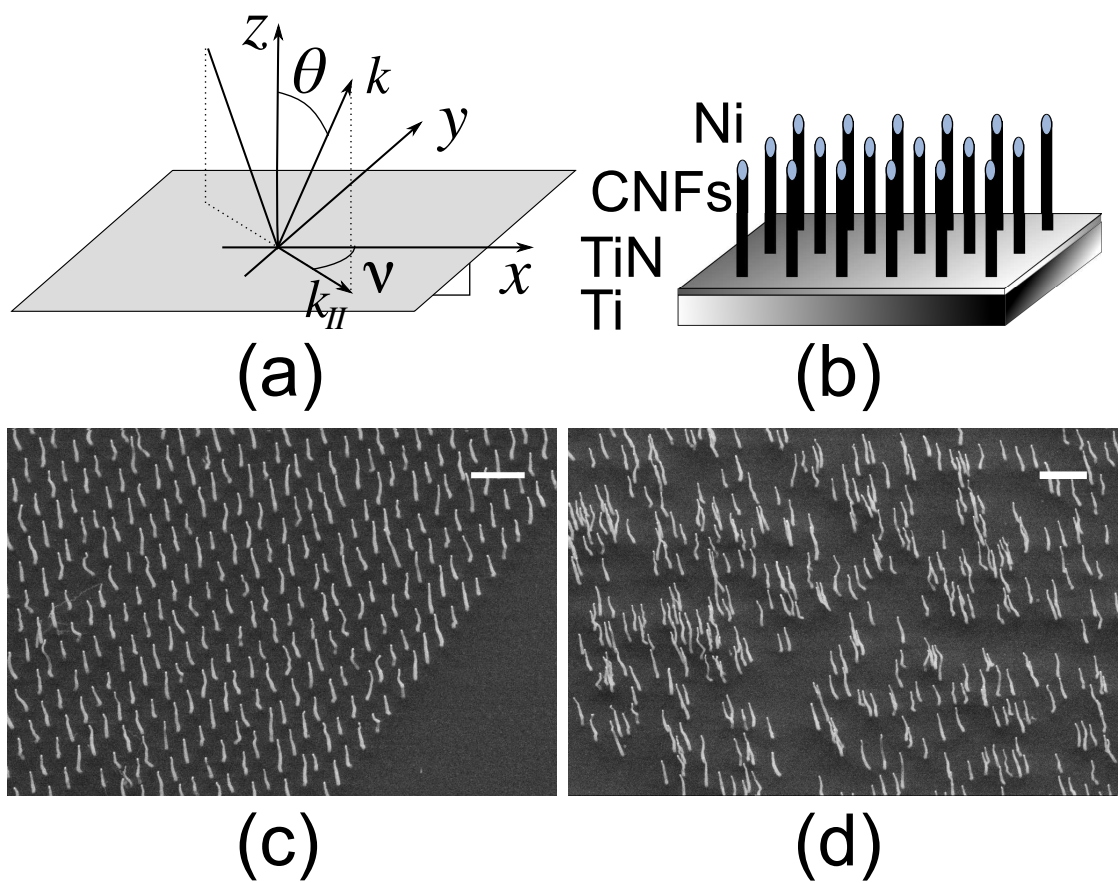}
    \caption{\label{fig:sample_images}(a) Ellipsometry setup with angle of incidence, $\theta$, and sample rotation angle, $\nu$, marked. (b) Sample illustration with layer structure. SEM micro graphs of (c) square lattice with $500$ nm lattice constant ($\sim 0.8$ \% filling fraction) and (d) random pattern with the same CNF filling factor as the sample in (c). Scale bars in (c) and (d) are 1 $\mu$m.}
\end{figure}

Titanium substrates coated with a $20$ nm titanium nitride layer were used as the underlying support. The size of the Ni catalyst dots was $50\times50\times30$ nm$^3$ (in $x\times y\times z$ directions respectively). The growth chamber was first pumped to $10^{-7}$ Torr before ramping the temperature by $100~^\circ$C/minute while ammonia was introduced to the chamber at $60$ sccm. The DC plasma was ignited at a pressure of $4$ Torr and a substrate temperature of $500~^\circ$C. This step is used to clean the substrate and activate the catalyst particles and is carried out for $2$ minutes with a plasma current of $4.3$ mA cm$^{-2}$. During the activation step the temperature was raised to $600~^\circ$C at a rate of $50~^\circ$C/min. At $600~^\circ$C, acetylene was introduced to the chamber at $15$ sccm and the plasma current reduced by $33$\%. The temperature was ramped to $700~^\circ$C with a ramping speed of $50~^\circ$C/min. Growth time was adjusted depending on the desired height of the CNFs. A height of approximately $1$ $\mu$m is reached after $20$ minutes. The CNFs were found to grow via a tip-growth mechanism with the Ni particle situated at the tip after growth. The samples used for the ellipsometry measurements were ca. $1$ $\mu$m long, $50$ nm diameter CNFs. An example of a square lattice and of a random pattern are shown in panels (c) and (d) of Figure~\ref{fig:sample_images} along with a schematic illustration of the PC (b) and a sketch illustrating the experimental geometry (a).

In reflection ellipsometry, the polarization change of light reflected off a sample surface is measured. This change depends on the optical properties and the structure of the sample, including any overlayers. The basic quantity measured is
\begin{equation}
    \rho = \frac{R_p}{R_s} = \tan \Psi e^{i \Delta},
    \label{eq:ellipsometric_parameters}
\end{equation}
where $R_p$ and $R_s$ are the complex-valued reflection coefficients for light polarized parallel ($p$) and perpendicular ($s$) to the plane of incidence. $\Psi$ and $\Delta$ are called the ellipsometric angles. The ratio of amplitude change for $p$- and $s$-polarization is given by $\tan \Psi$. The corresponding difference in phase changes is given by $\Delta$.

However, if a sample is anisotropic, it is generally not possible to fully characterize it with ordinary ellipsometry, which only takes into account the diagonal elements of the reflection Jones matrix \cite{saleh_fundamentals_2007}, as was done in an analysis of 2D Si PC by Hsieh et al. \cite{hsieh_optical_2004}. When the off-diagonal elements of the Jones matrix are non-zero, generalized ellipsometry is needed. This will ensure that conversion between $s$- and $p$-polarization is correctly taken into account. In generalized ellipsometry, three complex-valued parameters are defined as
\begin{eqnarray}
    \frac{R_{pp}}{R_{ss}} &=& \tan \Psi_{pp} e^{i \Delta_{pp}}\nonumber\\
    \frac{R_{ps}}{R_{pp}} &=& \tan \Psi_{ps} e^{i \Delta_{ps}}\label{eq:ellipsometric_parameters_jones}
\\
    \frac{R_{sp}}{R_{ss}} &=& \tan \Psi_{sp} e^{i \Delta_{sp}}\nonumber,
\end{eqnarray}
where $R_{pp}$ and $R_{ss}$ are the reflection coefficients for the $p$- and $s$-polarizations, respectively, and $R_{ps}$ and $R_{sp}$ represent the coupling from $p$- to $s$-polarization and from $s$- to $p$-polarization, respectively. In this study only $\Psi_{pp}$ data (denoted simply as $\Psi$ for convenience) are presented. This choice is based on the fact that $\Psi$ is the most fundamental parameter in generalized ellipsometry. It is also the easiest parameter to start analyzing and has the closest connection to PC structure. A more detailed ellipsometric analysis, considering a larger part of the Mueller matrix, will be presented elsewhere. From (\ref{eq:ellipsometric_parameters_jones}) we find that
\[ \tan \Psi = \left|\frac{R_{pp}}{R_{ss}}\right| \]
which thus represents the ratio between $R_{pp}$ and $R_{ss}$.

The ellipsometer used in this study is a dual-rotating-compensator ellipsometer from J.A. Woollam Co. Inc. The instrument provides all 16 elements of the reflection Mueller matrix from which the generalized ellipsometry parameters can be derived \cite{humlicek_polarized_2005}. The full Mueller matrix was measured at different angles of incidence, $\theta$, and different rotations of the sample, $\nu$ (see Figure~\ref{fig:sample_images}(a) for angle definitions).

\begin{figure}
    \includegraphics[width=.95\linewidth]{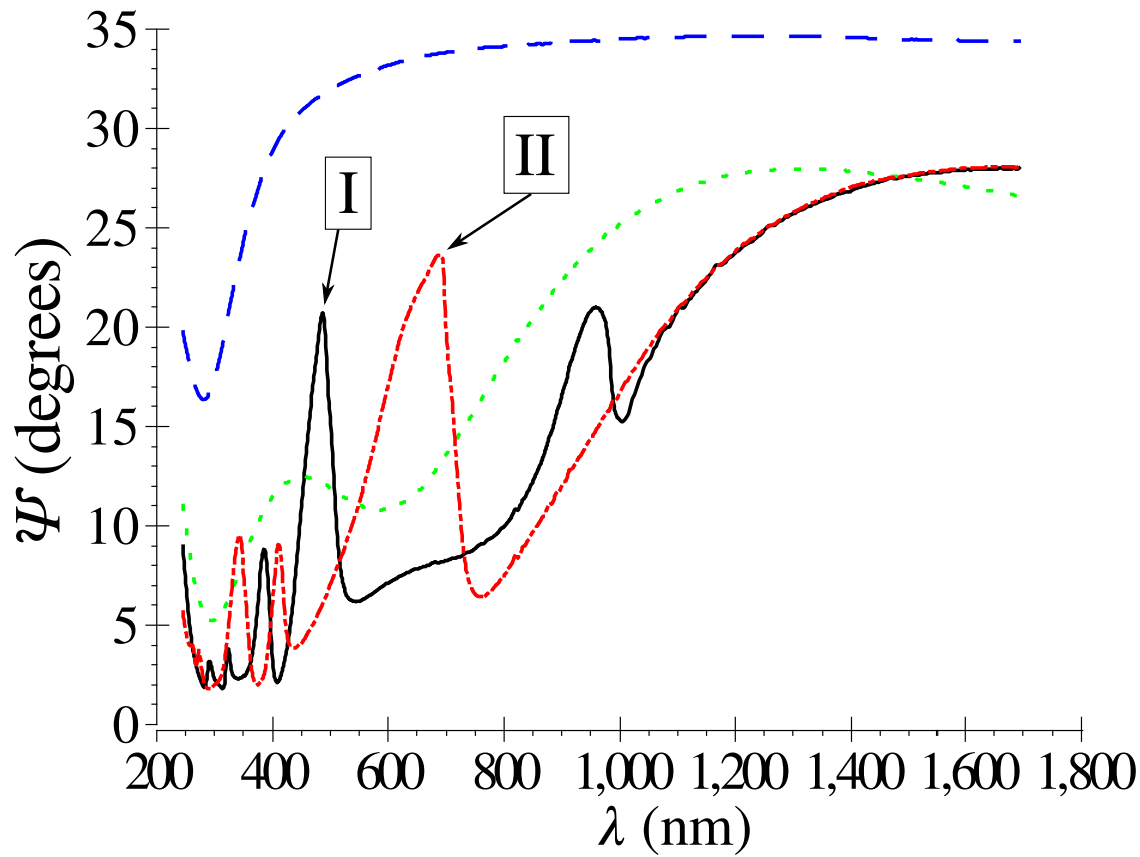}
    \caption{\label{fig:lattice_vs_random}Ellipsometric parameter $\Psi$, see (\ref{eq:ellipsometric_parameters}), plotted against wavelength for the square lattice with lattice constant $500$ nm, $\nu = 0^\circ$ (solid black) and $\nu = 45^\circ$ (dashed-dotted red), and for the random pattern (dotted green) with the same CNF filling factor. Also ‎$\Psi$ for the bare substrate is shown (dashed blue) as a reference. All curves are shown for angle of incidence $\theta = 75^\circ$. The second (first) peak that appears for the square lattice for $\nu = 0^\circ$ ($\nu = 45^\circ$) is marked with an arrow and denoted I (II).}
\end{figure}

For the 500 nm $\times$ 500 nm lattice, the ellipsometry parameter $\Psi$, measured at two sample orientations, $\nu = 0^\circ$ and $\nu = 45^\circ$ (where $\nu = 0$ correspond to the $x$-direction of the sample), is plotted against wavelength in Figure~\ref{fig:lattice_vs_random}. Measurements from the bare substrate and the random pattern are also shown for comparison. The data from the bare substrate show no pronounced features (except for the dip around $300$ nm which is due to the low refractive index of titanium at that wavelength \cite{palik}). The random pattern shows a fairly smooth variation in $\Psi$ but with a value that is significantly lower than for the bare substrate. This indicates that $p$-polarization is reflected much less strongly than $s$-polarization for the random pattern compared to the bare substrate. This effect can be related to the coupling of the incident $p$-polarized light to the surface plasmon polaritons (SPPs) \cite{maier_plasmonics_2007} supported by the metallic substrate. Note that this coupling is only possible due to the scattering of the incoming fields with the disordered CNFs at the substrate surface. 

Peaks appear in the $\Psi$ vs. $\lambda$ plot for the square and rectangular lattice structures as a consequence of their periodicity. Due to the regular lattice, the system support bound electromagnetic modes characterized by their optical band structure \cite{brillouin_wave_1953}. Since the CNFs are narrow (compared to the lattice constant) they represent a weak modulation of the metallic substrate surface. This implies that the first (lowest) band will not be much affected by the CNF modulation, remaining very similar to the dispersion relation of the SPPs supported by the bare substrate. However, at the Brillioun zone (BZ) edge a photonic band gap will open due to the presence of the CNF PC.

The group velocity in the PC is related to the modal dispersion via the relation
\[ \mathbf{v}_g(\mathbf{k}) = \nabla_{\mathbf{k}}\omega, \]
where $\mathbf{k}$ denotes the photonic Bloch wavevector in the PC, and $\omega$ the angular frequency. The resulting effective refractive index $n_\eff$ is direction dependent (a tensor) due to the anisotropic dispersion relation. In particular, at the BZ edge $\partial \omega/\partial k \rightarrow 0$, corresponding to a diverging $n_\eff$ in the substrate plane, which yields an impedance mis-match between the vacuum and the PC regions and causes reflection, mostly for the $p$-polarized light whose dispersion relation is more strongly affected by the PC. Therefore, we can expect sharp spectral features at wavelengths for which the dispersion relation of the modes supported by the structure yields $k$-vectors at the 2D BZ edge. If we consider the data shown in Figure~\ref{fig:lattice_vs_random} for the square $500 \times 500$ nm$^2$ lattice, the longest wavelength (smallest $k$) maxima for $\nu = 0^\circ$ and $\nu = 45^\circ$ appear at ca. $950$ nm and ca. $690$ nm, respectively. The corresponding wave vectors, $6.6$ $\mu$m$^{-1}$ and $9.2$ $\mu$m$^{-1}$, are simply related by $\cos(45^\circ) \simeq 0.71$ as one would expect from considering the geometry of the BZ.

It should be noted that it is the in-plane refractive index that diverges. This can, however, still give rise to an increase in reflection in the $z$-direction, see e.g. \cite{jackson_classical_1998}.

\begin{figure}
    \includegraphics[width=.95\linewidth]{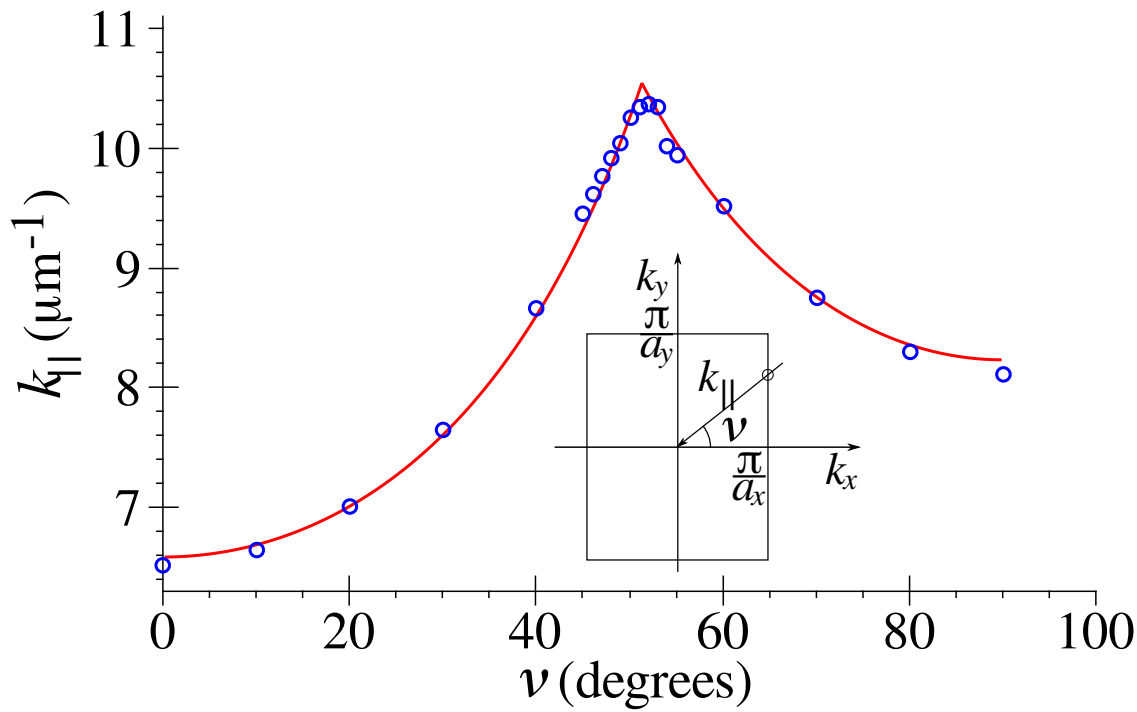}
    \caption{\label{fig:k_vs_nu}Experimental (circles) and calculated (curve) $k$-vector magnitude of the peak maximum as a function of the azimuthal angle, $\nu$, for the first peak (longest wavelength) in the $500 \times400$ nm$^2$ lattice, corresponding to the longest wavelength peak seen in Figure~\ref{fig:lattice_vs_random} for the square lattice.}
\end{figure}

A rectangular 2D PC has a reciprocal lattice that is also rectangular \cite{kittel_introduction_1996}, see inset in Figure~\ref{fig:k_vs_nu}. Here we consider rectangular lattices with CNF at positions $\mathbf{R}_{pq} = p~500~\mathrm{nm}~\hat{x} + q~400~\mathrm{nm}~\hat{y}$ where $p$ and $q$ are integers. From such geometry, we expect the first spectral maximum to correspond to an edge of the BZ that is either parallel with the $y$-axis (perpendicular to direction (10)) or with the $x$-axis (perpendicular to direction (01)). When the  azimuthal angle $\nu$ is varied, we expect the peak position to shift according to
\[ k_{\mathrm{peak}} = \left\{\begin{array}{l}
    \displaystyle \frac{k_{(10)}^{(0)}}{\cos \nu},~\tan \nu \leq \frac{500~\mathrm{nm}}{400~\mathrm{nm}}    \vspace{0.1cm}\\
    \displaystyle \frac{k_{(01)}^{(0)}}{\sin \nu},~\tan \nu > \frac{500~\mathrm{nm}}{400~\mathrm{nm}}.
    \end{array}\right. \]
Here $k_{(10)}^{(0)}$ is the wave vector of the first peak for a rotation angle $\nu$ such that the specular beam in the ellipsometer is in the $(10)$-direction. The parameter $k_{(01)}^{(0)}$ is similarly obtained for a specular beam in the $(01)$ direction. From the geometry of the BZ it follows that
\begin{equation}
    k_{(01)}^{(0)} = \frac{5}{4} k_{(10)}^{(0)}.
    \label{eq:k_relation}
\end{equation}
In Figure~\ref{fig:k_vs_nu}, the lowest wavelength peak is plotted together with the predicted $k_{\mathrm{peak}}$. Remarkably, the agreement between the peak position and this prediction is excellent.

\begin{figure}
    \includegraphics[width=0.95\linewidth]{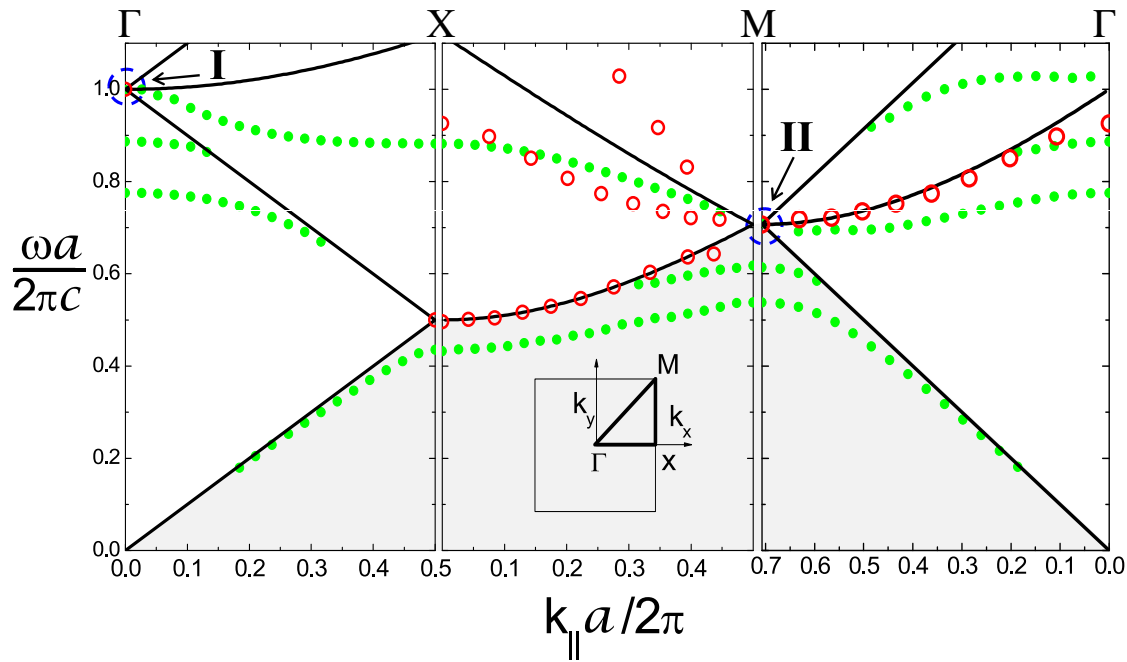}
    \caption{\label{fig:band_structure}Band structure of a 2D square PC consisting of dielectric pillars with radius of $0.05a$ and a refractive index of 4.1, calculated using finite difference time domain computations. The circles mark peak max of $\Psi$ for different rotations. The blue dashed circles I and II correspond to peaks I and II in Figure~\ref{fig:lattice_vs_random}. Inset shows the BZ of the square lattice with high symmetry points marked and the irreducible BZ highlighted. Band structure is filled green circles and solid black lines are light lines.}
\end{figure}

Next, we wish to understand the physical origin of the other peaks that appear in the spectra. The band structure is calculated using the finite difference time domain method \cite{bondeson_computational_2005}. Although the geometric structure of the PC is well defined (Figure~\ref{fig:sample_images}), the refractive index of the CNFs is not known. We therefore calculate the band structure for several different values of the refractive index and compare the results to the ellipsometric data. Best agreement is found for a choice of $n_\mathrm{CNF} = 4.1$ \footnote{This can only be considered a phenomenological model. For instance, the refractive index of the CNF cannot be expected to be constant over such a wide frequency range and no dissipative effects associated with the conductivity of the CNF have been included.}. The band structure for a 2D square lattice PC calculated using this value of $n_\mathrm{CNF}$ is shown in Figure~\ref{fig:band_structure}. Note that for PC with high-index pillars embedded in low-index material $s$-polarization in the ellipsometry setup is only weakly affected while the $p$-polarization in ellipsometry is strongly affected \cite{joannopoulos_photonic_2007}.

As can be seen in Figure~\ref{fig:band_structure}, for $p$-polarization, there are several regions of $(\omega,~k)$ space where the band structure is flat, leading to a divergence in the effective refractive index, $n_\eff$. The high effective refractive index causes strong reflection of the $p$-polarized light, increasing $R_p$ in (\ref{eq:ellipsometric_parameters}), and hence showing up as peaks in $\Psi(\lambda)$. Since the $s$-polarization is only weakly affected by the CNFs, we basically see only peaks and no dips in $\Psi$, as were observed for the Si PC reported by Hsieh et al. \cite{hsieh_optical_2004}.

To test if the detected peaks in $\Psi$ can be related to the band structure, $\Psi$ was recorded for an angle of incidence $\theta = 75^\circ$. The in-plane wave vector $k_\parallel$ was obtained as $k_\parallel = \sin\theta~k$ where $k=\omega/c$ is the free space wave vector. In Figure~\ref{fig:band_structure}, the experimental data are shown as empty red dots superimposed on the calculated band structure (solid green dots). The figure demonstrates how our experimental technique allows us to map the dispersion relation of the electromagnetic modes supported by the CNF PC. Note that the aim of this work is to test the capability of ellipsometry measurements to study guided modes in 2D PCs and that a more exhaustive reproduction of the theoretical bands is out of the scope of this paper. Our experimental results are in good agreement with theoretical data even in the XM axis, which is harder to probe experimentally. 

The theoretical bands within the shaded region in Figure~\ref{fig:band_structure} correspond to guided modes that lie below the lowest light line. These modes do not experience any radiation leakage while propagating along the PC structure, which implies that, in principle, they cannot be excited by ellipsometry measurements. Note that energy and momentum must be conserved in the scattering of the incident light with the sample. This explains why these modes remain hidden in the measurements along the $\Gamma$-X and M-$\Gamma$. Along the X-M direction, however, our experiments seem to reproduce the second lowest dispersion band (below the light line). The appearance of non-leaky modes supported by 2D PCs in reflection experiments has been recently reported by Paraire and co-workers \cite{paraire_investigation_2007}. In this work, this effect is linked to the presence of imperfections and to the finite size of the experimental sample.

The solid black lines in Figure~\ref{fig:band_structure} indicate the light lines. These correspond to the dispersion relation of grazing Bragg waves that propagates parallel to the sample surface. Note that, although these modes are not bounded to the PC structure, they play a relevant role in the scattering of the incoming light with the sample in our ellipsometry experiments \cite{joannopoulos_photonic_2007}. It should be noted that the guided PC modes tend to the light lines and eventually overlap at certain wavelengths within the BZ (see Figure~\ref{fig:band_structure}).

In Figure~\ref{fig:lattice_vs_random}, two large peaks are marked as I and II. These are depicted in Figure~\ref{fig:band_structure} as ellipses. They show up for the PC rotated in high symmetry directions, corresponding to directions with strong dispersion, as can also be seen in Figure~\ref{fig:band_structure}. Hence we conclude that ellipsometry also can be used to map symmetry directions of PCs when these are unknown.

In conclusion, there are no complete band gaps for any polarization in our samples due to the low filling factors of the PC under study. However, we still see strong effects at positions where the bands flatten out. The measured effects are much stronger than one might initially expect but they are in agreement with earlier computational results \cite{rehammar_nanowire-based_2008}.

Ellipsometry is a non-destructive, well-established technique for characterizing thin films. In this paper we have shown that it is possible to use ellipsometry to characterize 2D PC slabs and determine their band structure. It is also possible to easily identify symmetry directions and lattice parameters.

\ack
RR, JK and EEBC acknowledge funding from the SSF (Swedish Foundation for Strategic Research). RR, JK, EEBC, RM, HA acknowledge support from the Knut and Alice Wallenberg Foundation. We thank Peter Apell for stimulating discussions. EEBC acknowledges support from the WCU programme through KOSEF funded by MEST (R31-2008-000-10057-0).

\section*{References}
\bibliography{bibliography}{}
\bibliographystyle{unsrt}

\end{document}